\documentclass{relaxed_system_lab}
\usepackage[utf8]{inputenc}
\usepackage[T1]{fontenc}
\usepackage{charter}
\usepackage{amsmath,amssymb,amsfonts}
\usepackage{array}
\usepackage{makecell}
\usepackage{siunitx}
\usepackage{xspace}
\usepackage{longtable}
\usepackage{fontawesome5}
\usepackage[linesnumbered,ruled,vlined]{algorithm2e}

\newcolumntype{L}[1]{>{\raggedright\arraybackslash}p{#1}}

\newcommand{\sys}{\textsc{AReaL-TIK}\xspace}
\newtcolorbox{insightbox}{colback=blue!4!white,colframe=blue!45!black,
  boxrule=0.5pt,arc=1pt,left=5pt,right=5pt,top=4pt,bottom=4pt,
  before skip=7pt,after skip=5pt,
  fontupper={\bfseries\raggedright\hyphenpenalty=10000\relax}}

\AtBeginEnvironment{figure}{\setlength{\parskip}{0pt}}
\AtBeginEnvironment{figure*}{\setlength{\parskip}{0pt}}
\SetAlFnt{\normalsize}
\SetAlCapFnt{\normalsize}
\SetAlCapNameFnt{\normalsize}
\SetAlgoNlRelativeSize{0}
\hypersetup{colorlinks=true,linkcolor=labblue,citecolor=labblue,urlcolor=labblue}

\title{\sys: Stateful Agentic Optimization of Unified RL Kernels through an Optimization IR}
\author[1,*]{Ran Yan}
\author[1,*]{Youhe Jiang}
\author[2]{Jiayi Nie}
\author[1]{Wenshuang Li}
\author[3]{Yingqi Peng}
\author[4]{Taiyi Wang}
\author[3]{Tongkai Yang}
\author[1,3]{Binhang Yuan}
\affiliation{$^1$HKUST, $^2$University of Cambridge, $^3$Ant Group, $^4$Reflection AI}

\abstract{
Reinforcement learning (RL) post-training often uses distinct GPU kernels for rollout and policy update. In synchronous PPO and GRPO, numerical disagreement can perturb ratios between current token probabilities and those assigned during rollout. Recomputing rollout log-probabilities with the policy-update backend avoids this discrepancy but adds a forward pass. Bitwise-consistent unified kernels permit reuse when the policy snapshot and probability processing match the objective. Their optimization must preserve agreement across distinct execution regimes. We present \sys, an agentic framework starting from a hand-tuned, bitwise-consistent implementation. Its optimization intermediate representation (IR) organizes source-code search by linking implementations and modifications to numerical requirements, workload measurements, and derivation history. The agent coordinates changes and retains verified intermediates for further exploration; promotion requires passing correctness checks and improving aggregate latency within per-workload limits. Across 12 end-to-end training configurations on H20, \sys achieves $1.10\times$ average throughput relative to AReaL with log-probability recomputation, and the mean training-reward ratio rounds to $1.00\times$. Isolated-layer profiling yields $1.40\times$ average speedup in summed phase time across 15 model--GPU pairs. Operator-level evaluation covers correctness and performance for 10 operators on A100, H20, and H200, all passing the prescribed bitwise checks. Unified-attention search achieves $2.52\times$ speedup in summed workload latency over the starting implementation using 7M LLM tokens; ablations assess the contributions of retained evidence and branch exploration to search efficiency and attained performance.
}

\hypersetup{
  pdftitle={AReaL-TIK: Stateful Agentic Optimization of Unified RL Kernels through an Optimization IR},
  pdfauthor={Ran Yan, Youhe Jiang, Jiayi Nie, Wenshuang Li, Yingqi Peng, Taiyi Wang, Tongkai Yang, Binhang Yuan}
}

\begin{document}
\maketitle
\begingroup
\renewcommand{\thefootnote}{*}
\footnotetext[0]{Co-first authors: Ran Yan and Youhe Jiang. Corresponding author: Binhang Yuan (\href{mailto:biyuan@ust.hk}{biyuan@ust.hk}).}
\endgroup

\begin{center}
\vspace{-1.5em}
\href{https://github.com/areal-project/GameASG-Bench}{%
  \faGithub\,\texttt{\:Code: https://github.com/areal-project/AReaL-TIK}}
\vspace{0.5em}
\end{center}

\section{Introduction}

RL-based post-training has advanced the reasoning capabilities of large language models (LLMs)~\cite{openai2024o1,openai2026gpt56,guo2025deepseek,deepseek2026v4,qwen2024qwen2,anthropic2026fable}. RL systems generate responses during rollout and update model weights using those responses and their rewards~\cite{fu2025areal,sheng2024hybridflow,slime2025}. The two phases favor different GPU execution strategies, yet differences in their numerical computations can perturb the token-probability ratios used for training. Unified kernels seek to make corresponding forward computations agree bit for bit while retaining efficient execution in each phase. This paper studies how to automate the joint performance optimization of already consistent rollout and policy-update kernels while preserving bitwise-identical outputs for identical inputs and model weights.

Numerical consistency matters because the RL objective compares probabilities computed in the two phases. In standard synchronous Proximal Policy Optimization (PPO)~\cite{schulman2017proximal} and Group Relative Policy Optimization (GRPO)~\cite{shao2024deepseekmath}, each generated token's probability under the current model is divided by the fixed probability assigned by the \textit{behavior policy}, the model version that generated the response. Both probabilities condition on the same prompt and preceding response tokens. The logarithms of the denominator probabilities, conventionally called \textit{old-policy log-probabilities}, remain fixed while the model is updated on those responses. Before the first optimizer step, identical weights and probability processing should yield a ratio of one.

Separately optimized backends can violate this equality before any policy change. Policy-update attention commonly uses FlashAttention~\cite{dao2022flashattention,dao2023flashattention2,shah2024flashattention3,shah2025flashattention4}, whereas rollout engines such as SGLang~\cite{zheng2024sglang} and vLLM~\cite{kwon2023efficient} use specialized prefill and decode kernels from FlashInfer~\cite{ye2024flashinfer}. Differences in reduction order, precision conversion, and approximate instructions can change logits (unnormalized next-token scores) and token probabilities. Prior work shows that this Training--Inference Mismatch (TIM) can cause training collapse in controlled diagnostic experiments~\cite{zhong2026diagnosing}, intensify during optimization~\cite{zhang2026beyond}, and disproportionately affect low-probability tokens~\cite{wang2025taming}. Precision choices also matter: \texttt{BF16} rounding is a principal source of mismatch, and \texttt{FP16} can improve numerical agreement~\cite{qi2025defeating}.

\begin{figure}[!htbp]
    \centering
    \includegraphics[width=0.618\textwidth]{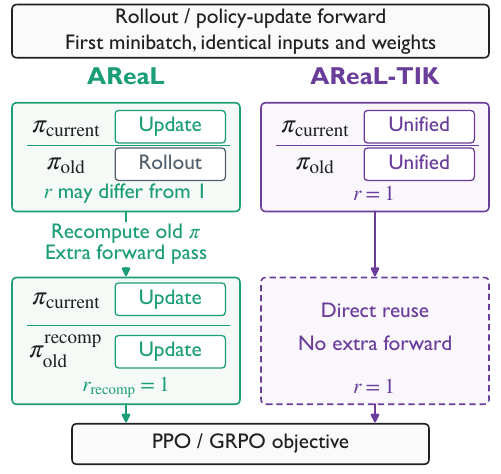}

    \caption{Token-probability ratios in the first minibatch, before any optimizer step, with identical inputs, weights, and probability processing. $\pi_{\mathrm{old}}$ denotes a recorded response token's probability under its generating behavior policy, conditioned on the same prompt and preceding tokens. AReaL computes $\pi_{\mathrm{current}}$ and $\pi_{\mathrm{old}}$ using different kernels, so direct reuse may yield $r=\pi_{\mathrm{current}}/\pi_{\mathrm{old}}\ne1$. An additional policy-update forward pass recomputes the denominator as $\pi_{\mathrm{old}}^{\mathrm{recomp}}$, restoring $r_{\mathrm{recomp}}=1$. \sys reuses recorded rollout token log-probabilities from consistent kernels (dashed box), yielding $r=1$ without recomputation.}
    \label{fig:ppo-grpo-logp-workflow}
    
\end{figure}

The AReaL baseline studied here evaluates the token log-probabilities needed for the ratio's fixed denominator with an additional policy-update forward pass over each selected prompt--response sequence. Figure~\ref{fig:ppo-grpo-logp-workflow} illustrates how this removes execution-path disagreement at identical weights. Unified kernels can avoid the extra pass by reusing recorded token log-probabilities when they match the policy snapshot and probability processing required by the objective. This condition also bounds the benefit under asynchronous training: consistent kernels do not make different policy versions interchangeable. In decoupled PPO~\cite{fu2025areal}, a newer \textit{proximal policy} may supply the clipped ratio's denominator; behavior-policy probabilities cannot replace its evaluation. This proximal policy is distinct from the reference model used for KL regularization.

Improving unified kernels requires optimizing two execution regimes. The policy-update forward pass and rollout prefill use a \textit{full-sequence entry point}\footnote{An \textit{entry point} is a callable host-side function that dispatches an invocation to a GPU kernel variant and launches it.}, processing a complete prompt--response sequence or the prompt, respectively. A \textit{token-wise decode entry point} processes one response token at a time against the key--value (KV) cache. Attention illustrates two challenges:

\begin{itemize}[leftmargin=*]
    \item \textbf{Preserving numerical agreement during specialization.} Full-sequence execution favors compute-intensive matrix multiplications, whereas decode is typically memory-bandwidth-bound. Each path therefore benefits from different data movement and GPU schedules. However, changes to tiling, reduction order, or intermediate precision can alter output values. Optimization must improve these specialized paths while preserving their numerical agreement.
    \item \textbf{Exploring coupled changes with delayed benefits.} Shared numerical and layout choices couple the two entry points: a change that accelerates one can slow the other or require a corresponding modification. Useful improvements can also require several steps, with an intermediate implementation enabling a later change without immediately improving aggregate performance. Optimizing each path in isolation, or retaining only the current fastest implementation, can miss such improvements.
\end{itemize}

Existing agentic kernel optimizers generate and evaluate source-code variants~\cite{saba2026cutegen,lange2025aicudaengineer,zhang2025cudaforge,cudaagent2026}, and evolutionary systems retain program history and evaluation feedback~\cite{novikov2025alphaevolve,sharma2025openevolve,liu2026skydiscover}. For unified kernels, the search must connect this information to numerical dependencies and performance effects across both entry points. A candidate's aggregate score alone cannot identify which numerical requirement a change violated, which workload limited its benefit, or which intermediate implementation enables a follow-up optimization.

Our approach exploits the distinction between numerical agreement and execution scheduling: the paths can use different GPU schedules provided their computations preserve the required numerical behavior. We present \sys, an agentic framework that starts from a bitwise-consistent hand-tuned implementation and searches for faster implementations of both paths. Its optimization intermediate representation (IR) organizes search state by linking source-code versions and proposed modifications to numerical requirements, workload measurements, and parent--child history. The agent uses these links to choose both what to change and which implementation to modify.

To preserve agreement during specialization, the IR associates each implementation with its shared source regions and numerical requirements, including reduction order and precision-conversion points. The agent consults these records when proposing full-sequence, decode, or joint modifications. The evaluator then checks bitwise agreement between the entry points and numerical accuracy against external references. Failed attempts remain linked to their source changes as evidence for later decisions.

To explore coupled changes, the evaluator measures both paths and promotes a candidate only if it reduces the weighted sum of workload latencies while respecting every workload's slowdown limit. The IR links these measurements to implementations and follow-up opportunities, allowing the agent to retain verified alternatives. In our H200 attention search, this lets the agent continue from a buffer-removal change that is not promoted but enables a later improvement (\S\ref{subsec:attention-case-study}).

Our contributions are as follows:

\noindent\textbf{\underline{Contribution 1.}} We develop an optimization IR for joint source-code search across full-sequence and token-wise decode entry points. It links complete implementations, numerical requirements, workload evidence, and derivation history to guide the agent's choice of modifications and parents (\S\ref{sec:optimization-ir}).

\noindent\textbf{\underline{Contribution 2.}} We implement IR-guided search that separates verification and performance promotion from branch selection. The agent can retain verified intermediates for subsequent changes even when they do not replace the current best. The workflow covers attention, normalization, position encoding, routing, activation, and KV-cache kernels (\S\ref{sec:ir-guided-search}).

\noindent\textbf{\underline{Contribution 3.}} We evaluate search effectiveness, training throughput and reward, isolated-layer phase times, and 10 operators. H20 unified-attention search achieves $2.52\times$ speedup in summed workload latency over the shared starting implementation using 7M LLM tokens; training throughput averages $1.10\times$ that of AReaL with recomputation across 12 configurations. Ablations distinguish search efficiency from attained performance: withholding numerical requirements raises token use to 16M, whereas restricting search to the current best lowers speedup to $2.00\times$ (\S\ref{subsec:ablation}).

\section{Related Work}

\noindent\textbf{RL training infrastructure.}
AReaL~\cite{fu2025areal}, veRL~\cite{sheng2024hybridflow}, OpenRLHF~\cite{hu2024openrlhf}, SLiME~\cite{slime2025}, and DeepSpeed-Chat~\cite{yao2023deepspeed} coordinate distributed rollout generation and policy update. AReaL overlaps these phases through asynchronous execution, while veRL supports flexible RL dataflows and efficient model resharding between generation and training. Such frameworks commonly integrate rollout engines~\cite{kwon2023efficient,zheng2024sglang} with policy-update backends~\cite{zhao2023pytorch,shoeybi2019megatron}. \sys complements these optimizations by jointly optimizing full-sequence and token-wise decode kernels, preserving bitwise consistency while allowing distinct GPU schedules.

\noindent\textbf{Training--inference mismatch in RL.}
Studies of training--inference mismatch (TIM) examine \texttt{BF16} rounding~\cite{qi2025defeating}, the growth of mismatch during training~\cite{zhang2026beyond}, and errors affecting low-probability tokens~\cite{wang2025taming}. Proposed mitigations include importance sampling~\cite{yao2025rollout}, adaptive learning rates~\cite{zhang2026beyond}, and deterministic inference~\cite{zhang2025deterministic}. VeXact~\cite{zhong2026diagnosing} establishes a zero-mismatch diagnostic setting using unified implementations whose outputs are invariant to batch composition. \sys focuses on improving the performance of already bitwise-consistent rollout and policy-update kernels while preserving their numerical agreement.

\noindent\textbf{Agentic GPU kernel development.}
GPU libraries and programming systems provide building blocks for kernel implementation and optimization~\cite{cublas,chetlur2014cudnn,cutlass,cute,tillet2019triton,spector2024thunderkittens,ansel2024pytorch2}. KernelBench~\cite{ouyang2025kernelbench} evaluates LLM-generated GPU kernels. CuTeGen~\cite{saba2026cutegen}, AI CUDA Engineer~\cite{lange2025aicudaengineer}, CudaForge~\cite{zhang2025cudaforge}, and CUDA Agent~\cite{cudaagent2026} automate kernel generation and refinement. \sys targets coupled full-sequence and token-wise decode entry points, evaluating source-code changes against both bitwise consistency between the paths and a joint performance objective with per-workload slowdown limits.

\noindent\textbf{In-context learning and persistent agent memory.}
Algorithm Distillation~\cite{laskin2023algorithm} and Retrieval-Augmented Decision Transformer~\cite{schmied2026retrieval} use interaction histories for in-context adaptation; Reflexion~\cite{shinn2023reflexion} retains episodic feedback to guide subsequent attempts. AlphaEvolve~\cite{novikov2025alphaevolve} combines LLM-generated source-code modifications, evaluators, and an evolutionary program database. OpenEvolve~\cite{sharma2025openevolve} and SkyDiscover~\cite{liu2026skydiscover} also retain program history and evaluation feedback. \sys organizes retained search experience for coupled kernels in an optimization IR linking numerical requirements and workload measurements to source implementations, follow-up opportunities, and derivation history. These links guide modifications and selection of verified parents, including intermediates, without training another optimization policy.

\section{Motivating Coupled Kernel Optimization}
\label{sec:coupled-kernel-optimization}

Figure~\ref{fig:ppo-grpo-logp-workflow} illustrates how numerical differences between rollout and policy-update kernels can perturb the RL probability ratio even before the model weights change. We first formulate this discrepancy and the consistency requirement used to avoid it, then examine how preserving that requirement shapes joint kernel optimization.

\subsection{Numerical Inconsistency}
\label{subsec:numerical-inconsistency}
Comparing execution paths at a fixed policy snapshot separates numerical disagreement from genuine policy change.

\noindent\textbf{Notation.} Scalars and indices use italic symbols; vectors and tensors use $\mathbf{p}$, $\mathbf{y}$, and $\mathbf{z}$. Calligraphic symbols denote sets, model execution, and structured objects, including $\mathcal{W}$, $\mathcal{M}$, $\mathcal{I}$, and $\mathcal{S}$. Action and audit records use $a$ and $e$; aggregate latency uses $T$.

Let $\mathbf{p}$ be a prompt and $\mathbf{y}$ a recorded response. For $i=1,\ldots,|\mathbf{y}|$, response token $y_i$ is conditioned on the prefix $\mathbf{x}_i=\mathbf{p}\Vert\mathbf{y}_{<i}$, where $\Vert$ denotes sequence concatenation. To isolate numerical disagreement, we hold model weights, tokens, positions, and causal masking fixed. For a complete source-code implementation $\mathcal{I}$ containing both execution paths, let $\mathbf{z}_{\mathcal{I},i}^{\mathrm{upd}}$ and $\mathbf{z}_{\mathcal{I},i}^{\mathrm{roll}}$ denote the logit vectors that predict $y_i$ through policy update and rollout, respectively. The update path processes the complete trajectory under causal masking. During rollout, the final prompt-prefill logits predict $y_1$; subsequent predictions use decode after processing the preceding response tokens, with the KV cache representing the same prefix $\mathbf{x}_i$.

Applying the same deterministic logit processing and normalization yields token log-probabilities $\ell_{\mathcal{I},i}^{\mathrm{upd}}$ and $\ell_{\mathcal{I},i}^{\mathrm{roll}}$. Define their discrepancy as $\delta_{\mathcal{I},i}=\ell_{\mathcal{I},i}^{\mathrm{upd}}-\ell_{\mathcal{I},i}^{\mathrm{roll}}$. Directly reusing the rollout value gives the probability ratio $r_{\mathcal{I},i}^{\mathrm{reuse}}=\exp\!\bigl(\delta_{\mathcal{I},i}\bigr)$.
Before the first optimizer step in synchronous training, the intended values are $\delta_{\mathcal{I},i}=0$ and $r_{\mathcal{I},i}^{\mathrm{reuse}}=1$. Differences in numerical execution can produce a nonzero $\delta_{\mathcal{I},i}$, perturbing the ratio despite unchanged weights. Recomputing the denominator through the policy-update path at the same snapshot removes this discrepancy but incurs another forward pass, as shown in Figure~\ref{fig:ppo-grpo-logp-workflow}. After an update, or with different current- and behavior-policy snapshots, a ratio other than one can reflect genuine policy change; the fixed-snapshot comparison isolates numerical disagreement.

\noindent\textbf{Required agreement.}
Unified execution targets a stronger condition than equality of the recorded token's probability: its corresponding logit vectors must agree bit for bit, $\mathbf{z}_{\mathcal{I},i}^{\mathrm{upd}}\stackrel{\mathrm{bit}}{=}\mathbf{z}_{\mathcal{I},i}^{\mathrm{roll}}$.
Under the common probability computation, this condition gives $\delta_{\mathcal{I},i}=0$ and permits reuse when the recorded values match the policy snapshot required by the objective. The equality compares the two paths within $\mathcal{I}$; coordinated changes may alter both paths' outputs relative to a parent implementation. Numerical accuracy against external references is checked separately. Section~\ref{subsec:problem-formulation} extends this response-token requirement to complete model outputs and defines the joint performance objective.

\subsection{Key Observations for Coupled Optimization}
\label{subsec:coupled-observations}
Two observations explain how the search benefits from considering both execution paths and retaining useful intermediate implementations. Attention illustrates the corresponding operator-level requirement: outputs must agree at matching query positions for the same query data and causal KV prefix. Full-sequence execution processes many query rows concurrently during rollout prefill and policy-update forward, whereas decode processes one new query at a time. These paths favor different GPU execution strategies, but numerical dependencies and shared source code couple their optimization. We denote the query, key, and value tensors by $\mathbf{Q}$, $\mathbf{K}$, and $\mathbf{V}$, respectively. The product $\mathbf{Q}\mathbf{K}^{\top}$ computes attention scores, $\mathbf{P}$ denotes the resulting normalized attention weights, and $\mathbf{P}\mathbf{V}$ computes the attention output.

\begin{insightbox}
Insight 1: Numerical constraints couple specialized execution paths.
\end{insightbox}
An edit that changes arithmetic in one path may require a corresponding edit in the other. Changes to shared data layouts can also affect both paths. The following examples show a trade-off and a joint improvement.
\begin{itemize}[leftmargin=*]
    \item[$\circ$] \textit{A resource improvement in one path can increase work in the other.} Attention partitions the KV sequence into tiles. Their boundaries determine online-softmax rescaling, \texttt{FP32}-to-\texttt{BF16} conversion of attention weights, and $\mathbf{P}\mathbf{V}$ accumulation order. In one full-sequence kernel, reducing the tile from 128 to 64 keys lowered shared memory per thread block (CTA) from approximately 56\,KiB to 34\,KiB, raising the residency limit from four to six CTAs per streaming multiprocessor (SM). Within this design, preserving bitwise agreement requires decode to use the same 64-key partition. Decode therefore processes twice as many sequential KV blocks and softmax/$\mathbf{P}\mathbf{V}$ updates. More full-sequence residency comes at the cost of additional decode work; the latency effects must be measured for both paths.

    \item[$\circ$] \textit{A shared layout change can benefit both entry points.} While optimizing long-context decode, we found that the shared memory layout for transposed $\mathbf{V}$ decomposed each 32\,KiB tile into 32 Tensor Memory Accelerator (TMA) transfers of 1\,KiB. Changing its CuTe tiling order produced two operations of 16\,KiB and reduced the static TMA instructions in each kernel from 34 to 4. Both entry points construct their TMA descriptors and warp-group matrix multiply-accumulate (WGMMA) operands from this layout definition in the shared source code, so the modification also changed the full-sequence entry point. It passed all 12 bitwise-consistency tests between the two entry points and improved decode by up to $1.27\times$ and full-sequence execution by up to $1.09\times$, with average gains of $1.15\times$ and $1.06\times$, respectively. A decode-motivated source-code modification can therefore benefit policy update and rollout prefill when both entry points are changed coherently.
\end{itemize}

These examples motivate coordinated edits and measurements: resource changes alone do not establish a speedup, and a modification targeting one path can affect the other's correctness or performance.

\begin{insightbox}
Insight 2: Intermediate implementations can enable later gains.
\end{insightbox}
A change can enable a useful follow-up without immediately improving performance. Retaining its verified implementation lets the search pursue that opportunity, even while another implementation remains the current best.

In our H200 attention search, removing a temporary $\mathbf{K}$ staging buffer produces a verified implementation that does not replace the current best. Its revised layout nevertheless enables a subsequent optimization of TMA loading for transposed $\mathbf{V}$ and WGMMA computation of $\mathbf{P}\mathbf{V}$. The agent retains the intermediate implementation and the follow-up opportunity, then uses them to reach an implementation that satisfies the aggregate promotion criteria. Section~\ref{subsec:attention-case-study} traces the associated records and decisions.

These observations identify three requirements for the search: numerical requirements must inform candidate edits, workload measurements must expose effects on both paths, and implementation history must preserve opportunities enabled by earlier changes. Section~\ref{sec:optimization-ir} defines an IR that links this information; Section~\ref{sec:ir-guided-search} explains how the agent and evaluator use it to guide exploration and promotion.

\section{Optimization IR}
\label{sec:optimization-ir}

The observations above require the search to connect numerical constraints, performance measurements, and opportunities across implementation versions. The optimization IR provides a schema for search records and their relationships, linking each source-code version to the evidence and opportunities associated with it. We define the optimization problem, records of each attempt, and persistent state that makes them available for later decisions.

\subsection{Optimization Problem}

\label{subsec:problem-formulation}

We first define the implementation being optimized, the required output agreement, and the joint latency objective. These definitions specify what the IR must describe and what the evaluator must measure.

\noindent\textbf{Implementation and execution.}
Using the prompt $\mathbf{p}$, response $\mathbf{y}$, and concatenation operator defined in \S\ref{sec:coupled-kernel-optimization}, the complete trajectory is $\mathbf{p}\Vert\mathbf{y}$.
The implementation $\mathcal{I}$ introduced in \S\ref{subsec:numerical-inconsistency} contains the model's complete kernel source code: the entry points, shared helpers, shape- and architecture-specific kernel variants, and dispatch rules needed by policy update and rollout generation. Depending on the model architecture, $\mathcal{I}$ includes attention, normalization, position encoding, routing, KV-cache, and activation operators. Operators invoked during both policy update and rollout generation are unified within $\mathcal{I}$; rollout-only operators, such as KV-cache writes, are optimized by the same workflow but are not themselves subject to equality between policy-update and rollout outputs.
We use $\mathcal{M}^{\mathrm{upd}}_{\mathcal{I}}(\cdot)$ to denote the forward pass used during policy update over a complete prompt--response sequence, $\mathcal{M}^{\mathrm{pre}}_{\mathcal{I}}(\cdot)$ to denote rollout prefill over a prompt, and $\mathcal{M}^{\mathrm{dec}}_{\mathcal{I}}(\cdot\mid\cdot)$ to denote autoregressive rollout decode conditioned on the cached prefix. Each function returns a matrix with one logit vector per processed token position, in sequence order. That vector scores the next token: the last prompt position scores the first response token, and each response position scores the following token. Thus, the response-token vector $\mathbf{z}_{\mathcal{I},i}^{\mathrm{upd}}$ in \S\ref{subsec:numerical-inconsistency} is row $|\mathbf{p}|+i-1$ of the policy-update output, using one-based indexing. Policy-update execution and rollout prefill invoke the full-sequence entry point, whereas autoregressive rollout decode invokes the token-wise decode entry point.

\noindent\textbf{Consistency constraint.}
During policy update, the model computes logits over the complete prompt--response trajectory in a single invocation, $\mathcal{M}^{\mathrm{upd}}_{\mathcal{I}}(\mathbf{p} \Vert \mathbf{y})$. For comparison, we replay the same recorded sequence through the rollout path: prefill over the prompt, $\mathcal{M}^{\mathrm{pre}}_{\mathcal{I}}(\mathbf{p})$, followed by processing the recorded response tokens one at a time, $\mathcal{M}^{\mathrm{dec}}_{\mathcal{I}}(\mathbf{y}\mid\mathbf{p})$. The replay uses the fixed-snapshot conditions in \S\ref{subsec:numerical-inconsistency}, with the decode cache representing the corresponding prefix.
Here, $\Vert$ also denotes concatenation of output matrices along their token-position dimension. We extend the response-token consistency requirement in \S\ref{subsec:numerical-inconsistency} to every corresponding output position. At prompt positions, policy-update logits are compared with rollout-prefill logits; at response positions, with logits from autoregressive decode of the same response tokens. For every supported prompt--response pair $(\mathbf{p},\mathbf{y})$, we require

\begin{equation}
\label{eq:consistency}
\mathcal{M}^{\mathrm{upd}}_{\mathcal{I}}(\mathbf{p} \Vert \mathbf{y})
\stackrel{\mathrm{bit}}{=}
\mathcal{M}^{\mathrm{pre}}_{\mathcal{I}}(\mathbf{p})
\Vert
\mathcal{M}^{\mathrm{dec}}_{\mathcal{I}}(\mathbf{y}\mid\mathbf{p})
\end{equation}

The sequence-wide comparison includes the final response-position logits, although scoring the recorded response does not require them. Reuse remains subject to the policy-snapshot and probability-computation conditions in \S\ref{sec:coupled-kernel-optimization}. Constraint~\eqref{eq:consistency} specifies the target model-level equality. Search verification uses prescribed tests; the operator-level checks are reported in \S\ref{subsec:breakdown}.

\noindent\textbf{Joint optimization objective.}
Let $\mathcal{W}^{\mathrm{full}}$ and $\mathcal{W}^{\mathrm{dec}}$ be the full-sequence and token-wise decode workload sets. The former includes rollout prefill over prompts and policy-update forward passes over complete trajectories. Each workload identifies its entry point, so the two sets are disjoint; their union is $\mathcal{W}$. For each workload $w\in\mathcal{W}$, let $\mathrm{lat}(\mathcal{I},w)$ denote its latency under implementation $\mathcal{I}$ and $c_w\geq0$ its fixed scalar weight. The joint objective is the weighted sum

\begin{equation}
\label{eq:joint-performance}
T(\mathcal{I})=\sum_{w\in\mathcal{W}}c_w\,\mathrm{lat}(\mathcal{I},w).
\end{equation}

We seek an implementation $\mathcal{I}$ minimizing $T(\mathcal{I})$ subject to Constraint~\eqref{eq:consistency}, fixed per-workload latency limits, and the prescribed numerical-accuracy and other correctness requirements in the evaluation specification below. This kernel-search objective is not end-to-end training time; search efficiency measures the LLM tokens and candidate evaluations needed for a verified improvement.

\noindent\textbf{Evaluation specification.}
The immutable specification $\mathcal{S}$ fixes the target hardware, workload sets, correctness tests, objective weights, and a baseline latency and maximum permitted slowdown for every workload. These fixed criteria make candidate measurements comparable within a search. An implementation is \textit{verified} when compilation and execution succeed and all prescribed correctness checks pass. A verified implementation is \textit{admissible} when complete measurements also establish that every workload meets its latency limit. \textit{Promotion} replaces the current best only with an admissible implementation having strictly lower $T(\mathcal{I})$.

\subsection{Records and Relationships}
\label{subsec:ir-records}

The IR links attempted modifications to source implementations, motivations, and measured outcomes so later decisions can use the evidence behind aggregate scores. Implementation records identify lineage vertices, action records identify derivations, and audit records attach evidence. Opportunity and lineage-management records describe possible next steps and branch status. Five record types express these relationships:
\begin{itemize}[leftmargin=*]
    \item \textit{Optimization-opportunity record.} An opportunity proposes a search direction: an optimization mechanism, affected workloads or entry points, supporting evidence, and records of potentially affected implementations, without prescribing an exact source-code modification. Audit records supply correctness outcomes, workload dispatch, latency, compiler-reported resource usage, profiler measurements, and generated instructions.

    \item \textit{Implementation record.} This record stores one complete source-code implementation, its target GPU architecture, entry points, shared source-code regions, dispatch rules, and numerical requirements between the entry points.

    \item \textit{Action record.} An action specifies the exact source-code modification, linking its originating opportunity and preceding implementation records to the resulting implementation record. It identifies targeted workloads and kernel variants, modification scope, affected source-code regions, expected hardware effect, numerical risks, and the measurement that would contradict its rationale.

    \item \textit{Audit record.} An audit stores compilation and execution outcomes, correctness results, observed kernel dispatch, per-workload latencies, and compiler, profiler, and generated-instruction evidence. It records $T(\mathcal{I})$ given all required workload latencies; failed checks and incomplete measurements remain evidence without implying verification or promotion.

    \item \textit{Lineage-management record.} This record stores the agent's interpretation of the audit, branch status, and evidence-supported follow-up opportunities.
\end{itemize}

The agent retrieves and summarizes linked records to choose source-code modifications and parents.

\subsection{Persistent Search State}
\label{subsec:ir-state}

The search state collects the records and separates the best measured implementation from alternatives retained for exploration. After $t$ completed candidate audits and their branch decisions, the state is

\begin{equation}
\mathcal{Z}_t=\langle\mathcal{S},\mathcal{G}_t,\mathcal{F}_t,\mathcal{R}_t,\mathcal{O}_t,\mathcal{I}_t^{*},\kappa_t\rangle.
\end{equation}

\begin{itemize}[leftmargin=*]
    \item $\mathcal{S}$ is the immutable evaluation specification defined in \S\ref{subsec:problem-formulation}. It keeps correctness and performance criteria fixed across attempts.

    \item $\mathcal{G}_t$ is the directed derivation graph of implementation lineage through iteration $t$. Its vertices identify complete source-code implementations through their implementation records, and each derivation links the parent implementation or implementations to an offspring through the corresponding action record. Linked audit, opportunity, and lineage-management records preserve the evidence and decisions associated with each attempt, including failed offspring and previously explored opportunities.

    \item $\mathcal{F}_t$ is the active frontier, a set of verified implementations with agent-identified immediate follow-up opportunities.

    \item $\mathcal{R}_t$ is the set of verified implementations archived for possible reuse, with no immediate follow-up opportunity.

    \item $\mathcal{O}_t$ is the set of pending optimization opportunities. Each opportunity references one or more verified implementations to which it may apply and supporting audit evidence in $\mathcal{G}_t$, which may include failed attempts.

    \item The current-best implementation $\mathcal{I}_t^{*}$ minimizes recorded $T(\mathcal{I})$ among admissible implementations.

    \item $\kappa_t$ is the counter of consecutive attempts without improvement, defined as the number of consecutive audited offspring that have not improved the current-best implementation. The stopping threshold $\kappa_{\max}$ is normally configured between 20 and 50; it limits unproductive search rather than certifying convergence.
\end{itemize}

Records remain in the derivation history after their implementations leave the frontier. The workflow below specifies how each attempt updates this history and the implementations retained for exploration.

\section{IR-Guided Kernel Search}
\label{sec:ir-guided-search}

To use retained evidence during optimization, \sys alternates source-code modifications, evaluation, and branch decisions. This workflow updates the IR: the agent proposes edits and manages branches, while evaluator results determine verification and promotion. A worked attention example shows how these decisions enable a sequence of improvements.

\subsection{Workflow and Initialization}
\label{subsec:workflow-initialization}
\begin{figure}[!htb]
    \centering
    \begin{minipage}[t]{0.42\textwidth}
    \vspace{0pt}
    \centering
    \includegraphics[width=\linewidth]{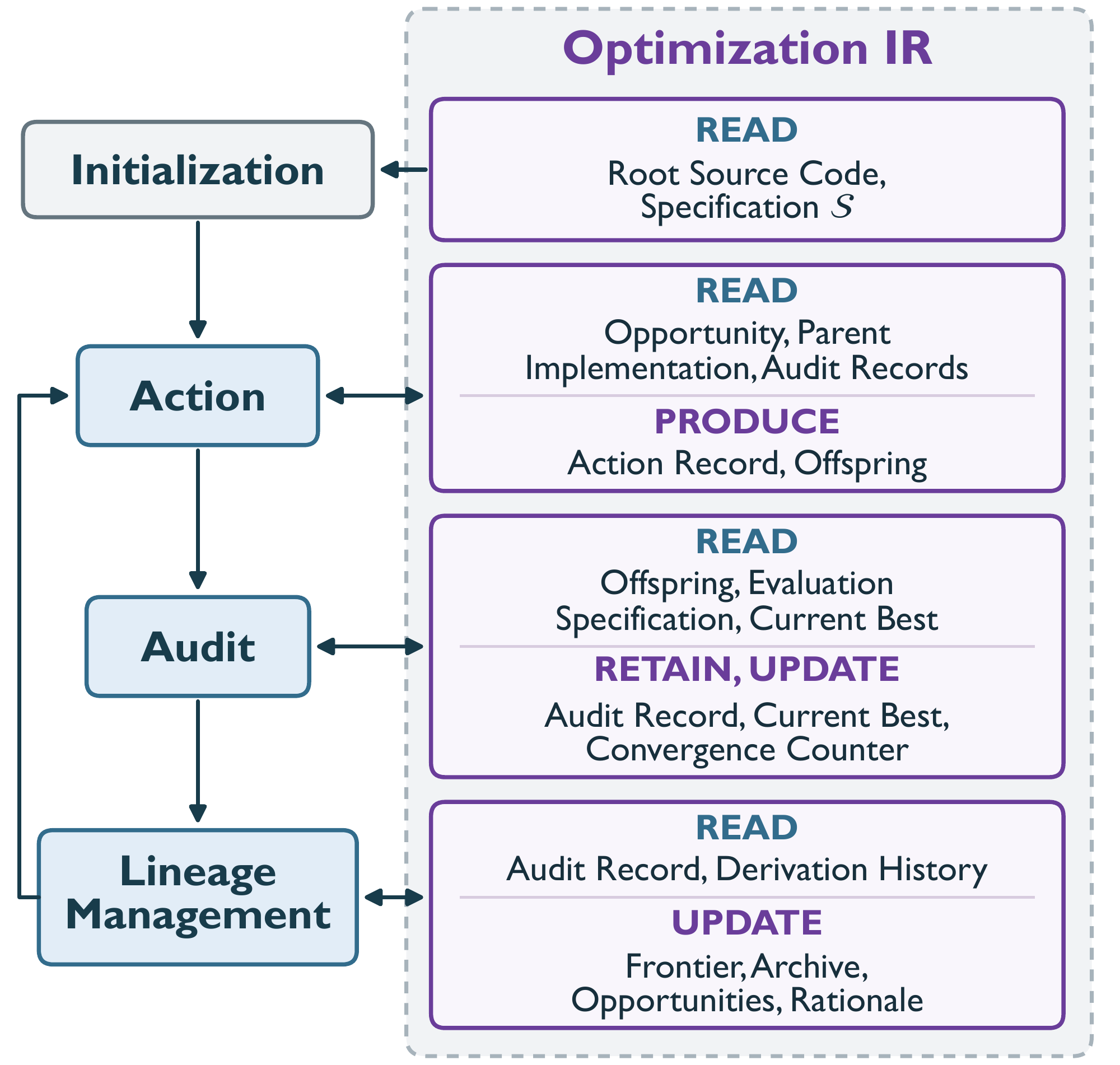}
    \caption{The agent loop and its IR records. \textit{Initialization} validates the root and creates the state; \textit{Action} produces an offspring from a selected opportunity and verified parent; \textit{Audit} records results and applies promotion criteria; \textit{Lineage Management} updates branches and opportunities. The figure's ``Convergence Counter'' counts consecutive audited attempts without improvement. Record types and state components are defined in \S\ref{subsec:ir-records} and \S\ref{subsec:ir-state}.}
    \label{fig:optimization-ir-comprehensive}
    \end{minipage}\hfill
    \begin{minipage}[t]{0.52\textwidth}
    \vspace{0pt}
\begin{algorithm}[H]
\normalsize
\DontPrintSemicolon
\caption{IR-Guided Unified RL Kernel Search}
\label{alg:ir-loop}
\KwIn{Bitwise-consistent starting implementation $\mathcal{I}_0$; specification $\mathcal{S}$; stopping threshold $\kappa_{\max}$}
\KwOut{Current best $\mathcal{I}_t^{*}$}
$e_0 \gets \textsc{Audit}(\mathcal{I}_0,\mathcal{S})$\;
\If{the root is unverified, incompletely measured, or inadmissible}{
    \textbf{stop without initializing the search}\;
}
$\mathcal{Z}_0 \gets \textsc{InitializeState}(\mathcal{S},\mathcal{I}_0,e_0)$; $t\gets0$\;
\While{$\mathcal{O}_t\neq\varnothing\ \text{and}\ \kappa_t<\kappa_{\max}$}{
    $(a_{t+1},\widehat{\mathcal{I}}_{t+1}) \gets \textsc{Action}(\mathcal{Z}_t)$\;
    $e_{t+1} \gets \textsc{Audit}(\widehat{\mathcal{I}}_{t+1},\mathcal{S})$\;
    $\mathcal{Z}_{t+1} \gets \textsc{RecordAudit}(\mathcal{Z}_t,a_{t+1},\widehat{\mathcal{I}}_{t+1},e_{t+1})$\;
    $\mathcal{Z}_{t+1} \gets \textsc{LineageManagement}(\mathcal{Z}_{t+1})$\;
    $t\gets t+1$\;
}
\textbf{return} $\mathcal{I}_t^{*}$\;
\end{algorithm}

In Algorithm~\ref{alg:ir-loop}, $t$ counts completed search attempts after initialization; $a_{t+1}$ records the selected parents and source-code modification, $\widehat{\mathcal{I}}_{t+1}$ is the resulting offspring, and $e_{t+1}$ contains its audit outcomes. The function $\textsc{RecordAudit}$ retains the action, offspring implementation record, and audit, including for failed attempts, and applies the promotion and counter updates. The loop stops when no pending opportunity remains or the non-improvement threshold is reached.
    \end{minipage}
\end{figure}

The workflow starts from a checked baseline so that every later attempt has a verified parent and comparable measurements. After \textit{Initialization}, it repeats \textit{Action}, \textit{Audit}, and \textit{Lineage Management}. Figure~\ref{fig:optimization-ir-comprehensive} shows the stages' interaction with the IR; Algorithm~\ref{alg:ir-loop} specifies state updates and the stopping rule.

\textit{Initialization} receives the evaluation specification $\mathcal{S}$ and a complete starting implementation $\mathcal{I}_0$, called the root, whose full-sequence and token-wise decode outputs already agree bit for bit; constructing this initial alignment is a prerequisite, not an optimization stage. The agent invokes the evaluator to verify and profile $\mathcal{I}_0$. The search starts only after compilation and execution succeed, every mandatory correctness check passes, all required workloads are measured, and the root satisfies the per-workload latency limits. \textit{Initialization} then creates $\mathcal{G}_0$, places $\mathcal{I}_0$ in $\mathcal{F}_0$, sets $\mathcal{R}_0=\varnothing$ and $\mathcal{I}_0^{*}=\mathcal{I}_0$, and initializes $\kappa_0=0$. The agent examines the root implementation and its audit evidence to derive the initial opportunities $\mathcal{O}_0$ used by the first \textit{Action}.

\subsection{Candidate Generation}
\label{subsec:candidate-generation}

Candidate generation turns a recorded opportunity into a concrete source-code change. In \textit{Action}, the agent interprets recorded results, selects a pending opportunity from $\mathcal{O}_t$, and chooses one or more verified parent implementations retained for exploration. An archived implementation can be reactivated when a new opportunity applies to it. Using the linked code and audit evidence, the agent chooses a source-code change expected to reduce $T(\mathcal{I})$ directly or enable an evidence-supported follow-up, while considering numerical and resource constraints. It records the modification's scope and rationale before applying it to produce a complete offspring.

\begin{figure*}[!htb]
    \centering
    \includegraphics[width=\textwidth]{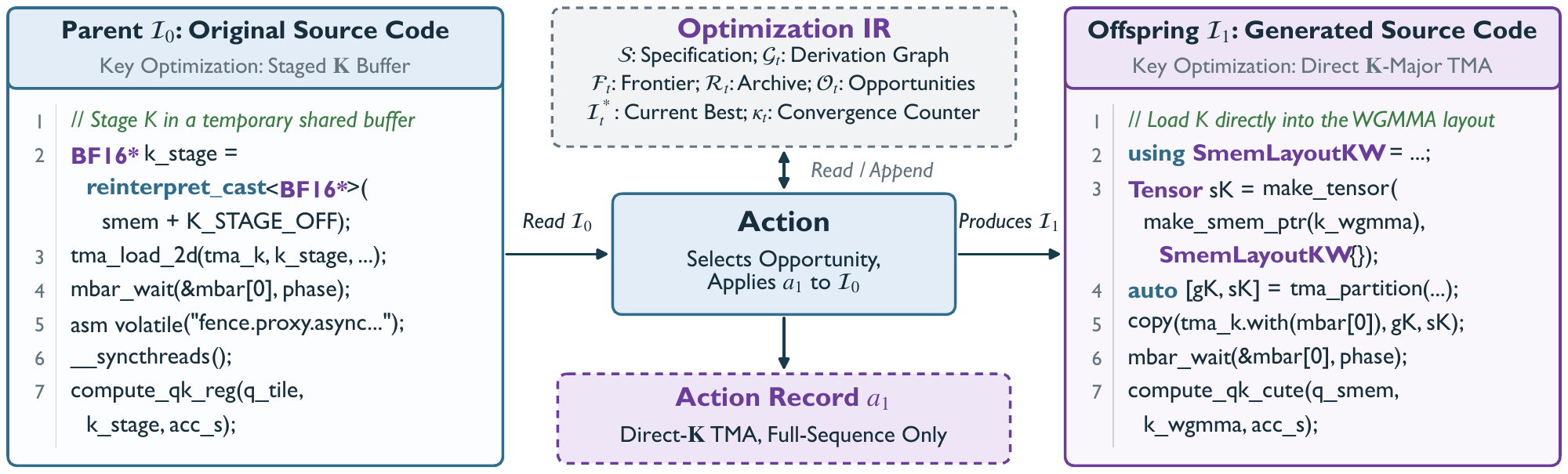}

    \caption{A source-code search step $a_1$: \textit{Action} reads root implementation $\mathcal{I}_0$ and its associated audit evidence, records TMA loading of $\mathbf{K}$ directly into WGMMA's required shared-memory layout, and produces offspring $\mathcal{I}_1$. The figure's ``Convergence Counter'' is $\kappa_t$, the count of consecutive audited attempts without improvement.}
    \label{fig:optimization-ir-transition}
    
\end{figure*}

The resulting action has one of three forms:

\begin{itemize}[leftmargin=*]
    \item[$\circ$] \textit{Full-sequence-only action.} The source code used by the policy-update forward pass and rollout prefill changes. The decode source code remains unchanged, providing the numerical counterpart for the bitwise-consistency check.

    \item[$\circ$] \textit{Decode-only action.} The token-wise rollout-decode source code changes. The full-sequence source code remains unchanged, providing the numerical counterpart for the bitwise-consistency check.

    \item[$\circ$] \textit{Joint action.} Both entry points or shared source code change. For promotion, a gain on one may offset the other's bounded slowdown only if $T(\mathcal{I})$ falls and all workload limits hold.
\end{itemize}

Every action creates, never overwrites, an implementation record, regardless of audit success.

\subsection{Verification and Promotion}
\label{subsec:verification-promotion}

Verified offspring can support further exploration; promotion also requires measured performance improvement. In \textit{Audit}, the agent invokes the evaluator to compile and execute the offspring under $\mathcal{S}$, check bitwise consistency between entry points and separately check numerical accuracy against external references, and measure the required workloads. An unverified offspring cannot enter the frontier or archive or become a parent. Its vertex remains in $\mathcal{G}_{t+1}$; its audit record retains failures and incomplete measurements.

The offspring $\widehat{\mathcal{I}}_{t+1}$ replaces the incumbent as $\mathcal{I}_{t+1}^{*}$ only if it is verified, fully measured, satisfies every per-workload latency limit in $\mathcal{S}$, and has lower aggregate latency: $T(\widehat{\mathcal{I}}_{t+1})<T(\mathcal{I}_t^{*})$. Promotion resets $\kappa_{t+1}$ to zero; otherwise, $\mathcal{I}_{t+1}^{*}=\mathcal{I}_t^{*}$ and $\kappa_{t+1}=\kappa_t+1$, including ties and failed audits. The agent selects branches separately from these updates.

For unified attention, recorded numerical requirements guide edits and audit-failure analysis across four risks: $(\textbf{i})$ the $\mathbf{Q}\mathbf{K}^{\top}$, $\mathbf{P}\mathbf{V}$, online-softmax, and final-normalization reduction orders; $(\textbf{ii})$ the precision and placement of scaling and masking; $(\textbf{iii})$ the exact exponential, reciprocal, and square-root instructions, including constants and whether subnormal values are flushed to zero; and $(\textbf{iv})$ the \texttt{BF16}--\texttt{FP32} conversion points and the order of accumulator rescaling, update, and normalization.

\subsection{Branch Exploration}
\label{subsec:branch-exploration}

\textit{Lineage Management} selects implementations to explore independently of promotion. After \textit{Audit} fixes $\mathcal{I}_{t+1}^{*}$, the agent reviews evidence, identifies follow-up opportunities, and records a branch decision without changing the incumbent:

\begin{itemize}[leftmargin=*]
    \item[$\circ$] \textsc{Continue} keeps a verified implementation active so that a long optimization path can be divided into short, separately verified actions. If an offspring fails, revisions must start from a retained verified parent, not that offspring.
    \item[$\circ$] \textsc{Merge} records an opportunity when the agent identifies compatible modifications on different branches; a later \textit{Action} constructs the combined implementation for \textit{Audit}.
    \item[$\circ$] \textsc{Archive} retains a verified implementation and its evidence without an immediate follow-up. The agent may reactivate it when new evidence supports further work.
    \item[$\circ$] \textsc{Abandon} stops work only after repeated retries remain unverified or the audit evidence falsifies the optimization direction; failed attempts remain as negative evidence.
\end{itemize}

\begin{figure}[!htb]
    \centering
    \includegraphics[width=0.618\textwidth]{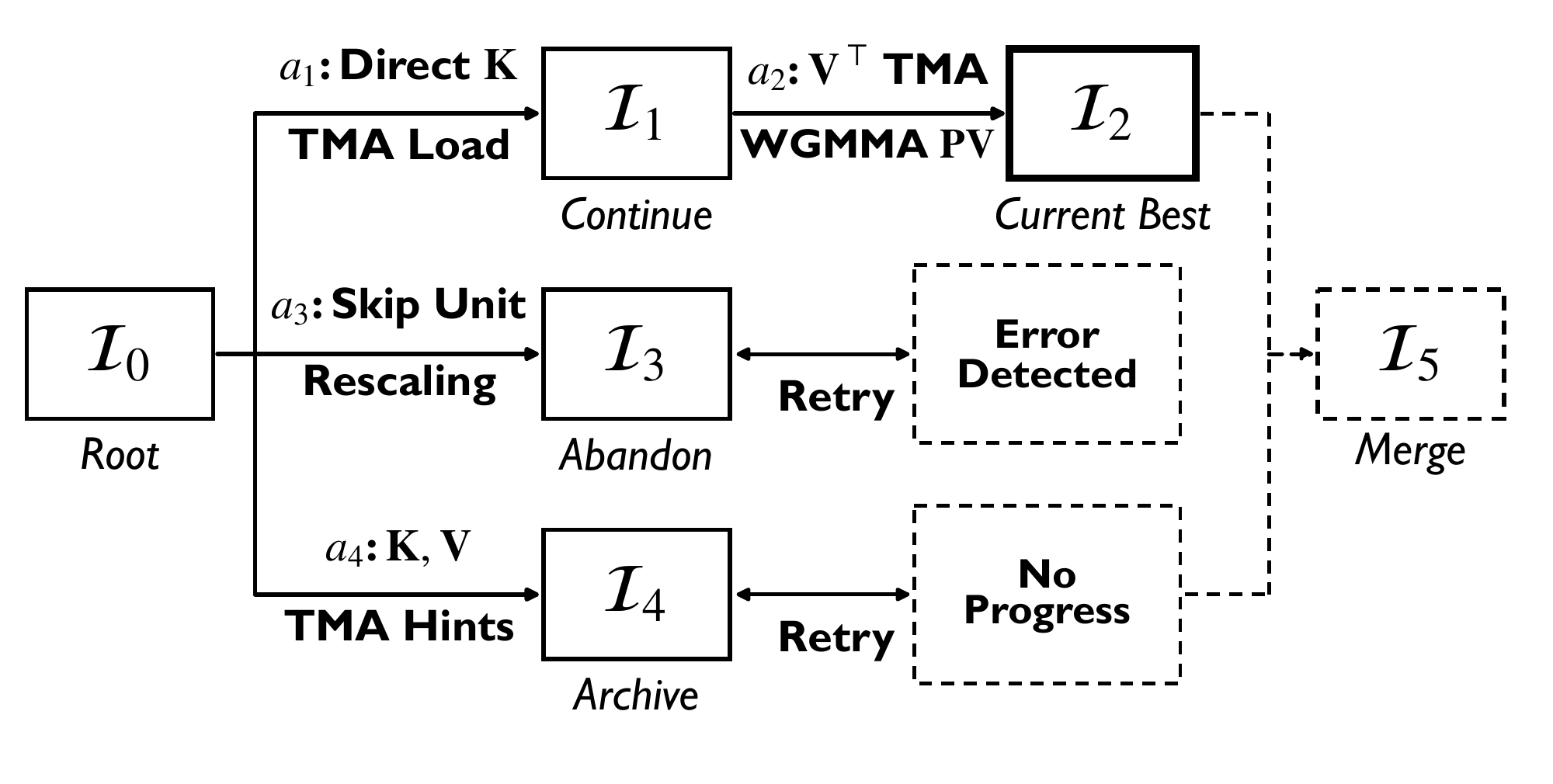}

    \caption{\textit{Lineage Management} for the unified-attention search rooted at $\mathcal{I}_0$. Solid arrows $a_1$--$a_4$ denote executed actions; dashed $\mathcal{I}_5$ is a proposed merge, neither constructed nor audited. ``Error Detected'' denotes failed bitwise checks; ``No Progress'' denotes no improvement of the aggregate objective. Retry arrows denote attempts from verified parent $\mathcal{I}_0$; failed $\mathcal{I}_3$ cannot be a parent.}
    \label{fig:optimization-ir-lineage}
    
\end{figure}

\subsection{Attention Case Study}
\label{subsec:attention-case-study}
The H200 attention search illustrates how IR records guide edits and parent choices. Here, $\mathcal{I}_0,\ldots,\mathcal{I}_4$ are constructed implementations, $\mathcal{I}_5$ is a proposed merge, and $a_1,\ldots,a_4$ are executed actions.

\noindent\textbf{From opportunity to offspring.}
Figure~\ref{fig:optimization-ir-transition} traces $a_1:\mathcal{I}_0\rightarrow\mathcal{I}_1$. In root implementation $\mathcal{I}_0$, the H200 full-sequence kernel stages $\mathbf{K}$ through a temporary 32\,KiB shared-memory buffer; an opportunity record links this source code to profiler evidence that the buffer limits block residency. The agent records and applies direct-$\mathbf{K}$ TMA loading into WGMMA's shared-memory layout, removing the buffer while leaving token-wise decode unchanged.

\noindent\textbf{Verification without promotion.} \textit{Audit} verifies $\mathcal{I}_1$: all mandatory correctness checks pass, shared memory falls from approximately 99 to 68\,KiB, and the residency limit rises from two to three CTAs per SM. $\mathcal{I}_1$ remains unpromoted under the aggregate objective and per-workload latency limits; its unchanged 1,024-token prefill latency of approximately 0.21\,ms cannot alone determine promotion.

\noindent\textbf{An intermediate implementation enables promotion.} In \autoref{fig:optimization-ir-lineage}, a follow-up opportunity identifies how $\mathcal{I}_1$'s layout supports TMA loading of transposed $\mathbf{V}$ and WGMMA computation of $\mathbf{P}\mathbf{V}$. \textit{Lineage Management} assigns \textsc{Continue}; \textit{Action} selects verified $\mathcal{I}_1$ rather than incumbent $\mathcal{I}_0$ for $a_2:\mathcal{I}_1\rightarrow\mathcal{I}_2$. $\mathcal{I}_2$ meets the promotion criteria, reducing 1,024-token prefill latency from approximately 0.21 to 0.10\,ms.

\noindent\textbf{Rejected and archived alternatives.} On another branch, $a_3$ produces $\mathcal{I}_3$ by skipping accumulator rescaling when the scale factor is one. Repeated variants fail bitwise checks because skipping the instruction changes whether subnormal values are flushed to zero, leading to \textsc{Abandon}; action and audit records retain the attempted simplification and rejection rationale. Action $a_4$ adds TMA cache hints for $\mathbf{K}$ and $\mathbf{V}$ to produce $\mathcal{I}_4$, which passes all 12 bitwise tests but reduces aggregate performance by 0.08\%, receiving \textsc{Archive}. Dashed $\mathcal{I}_5$ proposes combining this archived modification with $\mathcal{I}_2$; it has not been constructed or audited.

\section{Evaluation}

We evaluate optimized-kernel performance at the training, layer, and operator levels (Q1--Q3), compare search systems (Q4), and ablate \sys's components (Q5). Q1, Q4, and Q5 use NVIDIA H20 GPUs; Q2 and Q3 also include A100 and H200 GPUs.

\FloatBarrier
\subsection{Q1: End-to-End PPO and GRPO Training}
\label{subsec:end-to-end}

We train Qwen2.5-1.5B and Qwen3-4B with PPO and GRPO on GSM8K~\cite{openai2021gsm8k} at configured data-staleness values $s\in\{0,2,4\}$. Here, $s=0$ is synchronous and on-policy; $s=2$ and $s=4$ permit asynchronous, off-policy execution. For each configuration, AReaL and \sys use identical training settings for 300 policy updates, with reward and throughput averaged across three random seeds.

Under the same AReaL orchestration, the baseline combines \texttt{SGLang} rollout and \texttt{FSDP} policy-update kernels and recomputes token log-probabilities for the clipped ratio's fixed denominator; \sys uses unified kernels and reuses recorded rollout values. Numerical consistency alone does not make behavior-policy probabilities interchangeable with those of a distinct proximal-policy snapshot.

At each update, we report mean task reward over 256 selected trajectories, actor policy loss, and pre-clipping actor gradient norm. End-to-end throughput is completed training-response tokens divided by elapsed time from the first rollout dispatch through completion of in-flight work after update 300, including rollout, policy update, scheduled evaluation, checkpointing, synchronization, and queueing. We also estimate AReaL throughput without recomputation. Reported reward and throughput ratios compare \sys with AReaL and are averaged arithmetically across 12 configurations. Training hyperparameters appear in Appendix~\ref{app:training-hyperparameters}.

\begin{figure}[!htbp]
    \centering
    \begin{minipage}[t]{0.485\textwidth}
    \vspace{0pt}
    \centering
    \includegraphics[width=\linewidth]{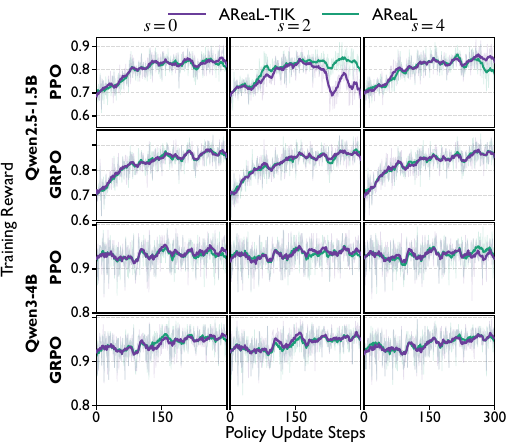}

    \caption{Training reward over 300 updates. Rows distinguish model and algorithm; columns show staleness $s\in\{0,2,4\}$. Faint lines average per-update rewards across three seeds, each with 256 trajectories selected for the update; opaque lines show centered 15-update moving averages.}
    \label{fig:training-reward-curves}
    \end{minipage}\hfill
    \begin{minipage}[t]{0.485\textwidth}
    \vspace{0pt}
    \centering
    \includegraphics[width=\linewidth]{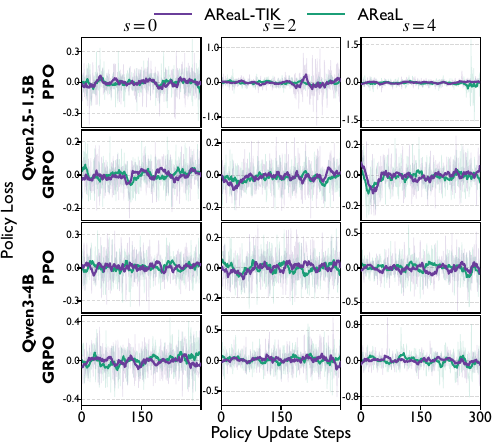}

    \caption{Actor policy loss ($10^{-3}$ units) for the configurations in \autoref{fig:training-reward-curves}. Faint lines show per-update values; opaque lines show centered 15-update moving averages.}
    \label{fig:training-policy-loss-curves}
    \end{minipage}
\end{figure}

\begin{figure}[!htbp]
    \centering
    \begin{minipage}[t]{0.485\textwidth}
    \vspace{0pt}
    \centering
    \includegraphics[width=\linewidth]{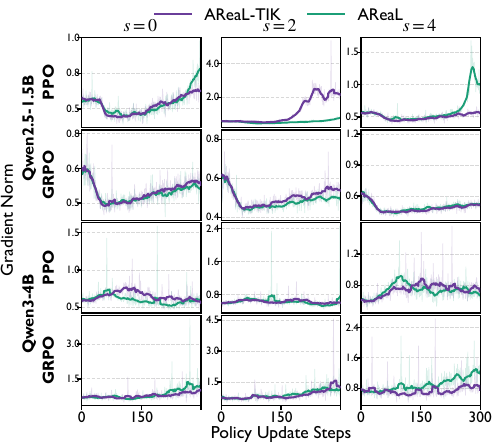}

    \caption{Pre-clipping actor gradient norm for the configurations in \autoref{fig:training-reward-curves}. Faint lines show per-update values; opaque lines show centered 15-update moving averages.}
    \label{fig:training-gradient-norm-curves}
    \end{minipage}\hfill
    \begin{minipage}[t]{0.485\textwidth}
    \vspace{0pt}
    \centering
    \includegraphics[width=\linewidth]{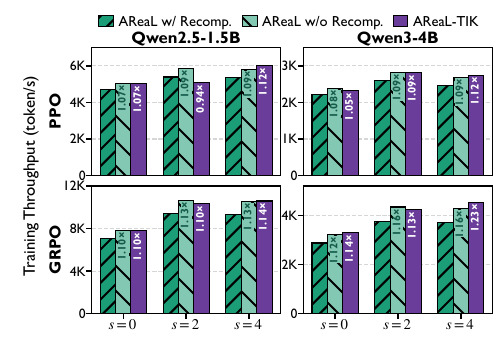}

    \caption{End-to-end training-response throughput over 300 policy updates at $s\in\{0,2,4\}$, averaged across three seeds. \sys and AReaL with recomputation are measured; AReaL without recomputation is estimated.}
    \label{fig:training-throughput}
    \end{minipage}
\end{figure}

\noindent\textbf{End-to-end results.} The average training-reward ratio of \sys to AReaL rounds to $1.00\times$ across 12 configurations, using each configuration's mean reward over all 300 updates (\autoref{fig:training-reward-curves}). GRPO and Qwen3-4B PPO range from $0.999\times$ to $1.003\times$, but Qwen2.5-1.5B PPO at $s=2$ reaches $0.955\times$. Actor policy loss and pre-clipping gradient norm (\autoref{fig:training-policy-loss-curves}, \autoref{fig:training-gradient-norm-curves}) diagnose optimization, not task performance; lower values need not indicate better training.

\noindent\textbf{Training throughput.} \sys averages $1.10\times$ AReaL's throughput with recomputation, improving 11 of 12 configurations; GRPO and PPO average $1.14\times$ and $1.07\times$ (\autoref{fig:training-throughput}). Without recomputation, AReaL's estimated throughput is $1.11\times$ the measured baseline. Measured gains combine unified kernels and rollout-log-probability reuse; Q4 instead measures search gains over a shared bitwise-consistent implementation.

\FloatBarrier
\subsection{Q2: Isolated-Layer Performance Across Models}
\label{subsec:across-models}

We profile isolated transformer layers from two dense models, Qwen3-4B and Qwen3-8B, and three MoE models, Mixtral-8$\times$7B, Qwen3-30B-A3B, and Qwen3-235B-A22B. We use batch size 1, a 1,024-token prefill, and decode lengths of 128, 1,024, 4,096, and 8,192 tokens. Attention shares the implementation in \S\ref{subsec:breakdown}, with model-specific query, key, and value tensor dimensions.

\begin{figure*}[!htb]
    \centering
    \includegraphics[width=\textwidth]{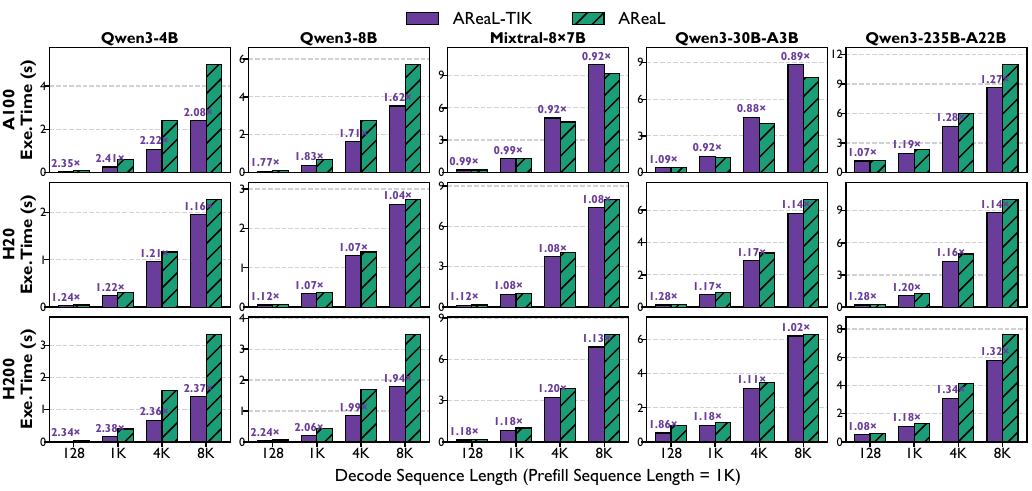}

    \caption{Aggregate isolated-layer phase time, including AReaL's recomputation forward pass. Rows show A100, H20, and H200; columns show the five models. Panels use independent vertical scales, decode lengths of 128, 1K, 4K, and 8K, and 1K prefill. Labels report the ratio of AReaL's summed phase time to \sys's.}
    \label{fig:per-layer-execution-times}
    
\end{figure*}

Per-layer time sums prompt prefill once, token-wise decode over the complete response, and policy-update forward and backward each once over the complete prompt--response trajectory. \sys uses unified kernels; AReaL uses \texttt{SGLang} rollout and \texttt{Megatron} policy-update kernels, plus a forward pass recomputing the objective's fixed-denominator token log-probabilities. This sum is not wall-clock training throughput and does not account for phase overlap.

\noindent\textbf{Layer performance.} Speedups average $1.40\times$ across 15 GPU--model pairs, with gains in 13 (\autoref{fig:per-layer-execution-times}). Qwen3-4B and Qwen3-8B average $2.10\times$ on A100 and H200; the five H20 models average $1.15\times$. Gains combine kernel-execution differences and avoided recomputation.

\begin{figure*}[!htb]
    \centering
    \includegraphics[width=\textwidth]{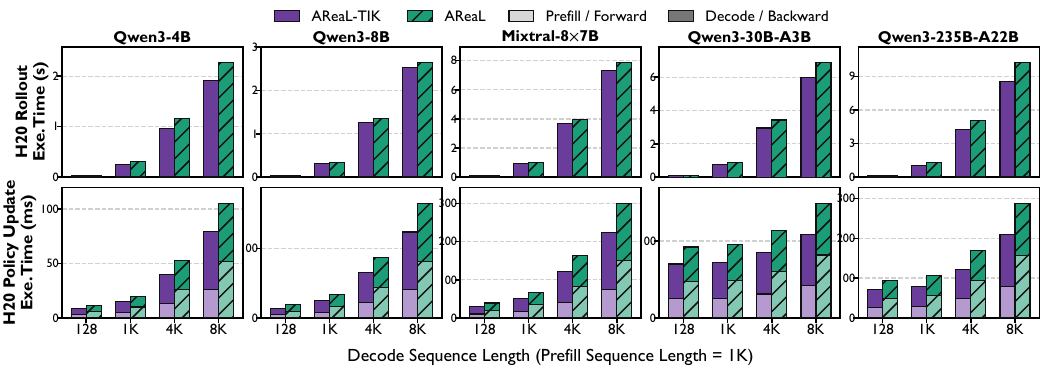}

    \caption{H20 per-layer phases: batch size 1, 1K prefill. Each response length compares \sys (purple) with AReaL (green, hatched). The top row stacks prefill and cumulative decode; the bottom stacks measured policy-update forward and backward times. AReaL's forward time also includes the extra pass for denominator token log-probabilities.}
    \label{fig:h20-layerwise-breakdown}
    
\end{figure*}

\noindent\textbf{H20 phase breakdown.} \sys reduces cumulative rollout-decode time in all 20 configurations (\autoref{fig:h20-layerwise-breakdown}). Avoided fixed-denominator recomputation lowers policy-update forward totals; backward times are comparable. With 1K prompts, prefill times are similar and contribute little.

\FloatBarrier
\subsection{Q3: Per-Operator Performance}
\label{subsec:breakdown}

Prefill workloads use 1K, 4K, 8K, and 16K tokens at batch size 1; decode combines lengths of 128, 1K, 4K, and 8K with batch sizes of 1, 4, 8, and 16. Attention baselines are FlashAttention-2 on A100, FlashAttention-3 on H20 and H200, and the fastest qualified FlashInfer configuration for each decode workload. The other nine operators use \texttt{nn.Linear}/\texttt{ATen-cuBLAS} for dense linear layers and the applicable \texttt{Megatron-Core} or \texttt{SGLang} implementations otherwise. Speedup is baseline latency divided by \sys latency; values below one indicate slower execution.

\noindent\textbf{Correctness checks.} Under Constraint~\eqref{eq:consistency}, we compare full-sequence and token-wise outputs for identical inputs per GPU and verify exact KV-cache post-write states (\autoref{tab:operator-correctness}). Attention passes 64 prompt--response cases and 36 batched-versus-single decode comparisons on A100, and 16 prompt--response cases with 128-token responses on H200. H20 passes entry-point consistency checks, 32 bitwise-preservation checks, and CUDA-graph checks with varying inputs.

\begin{table}[!htbp]
    \centering
    \caption{Prescribed bitwise checks for operator outputs and KV-cache post-write states. Checkmarks indicate passing the tests within each GPU's tested domain, not arbitrary-shape coverage or bitwise equality with external libraries.}
    \label{tab:operator-correctness}
    
    \normalsize
    \setlength{\tabcolsep}{4pt}
    \begin{tabular}{@{}L{4.35cm}|ccc@{}}
        \toprule
        \textbf{Operator} & \textbf{A100} & \textbf{H20} & \textbf{H200} \\
        \midrule
        Attention & $\checkmark$ & $\checkmark$ & $\checkmark$ \\
        Generic dense FFN & $\checkmark$ & $\checkmark$ & $\checkmark$ \\
        MoE router & $\checkmark$ & $\checkmark$ & $\checkmark$ \\
        Hidden RMSNorm & $\checkmark$ & $\checkmark$ & $\checkmark$ \\
        $\mathbf{Q}$ RMSNorm & $\checkmark$ & $\checkmark$ & $\checkmark$ \\
        $\mathbf{K}$ RMSNorm & $\checkmark$ & $\checkmark$ & $\checkmark$ \\
        Fused residual + RMSNorm & $\checkmark$ & $\checkmark$ & $\checkmark$ \\
        NeoX RoPE ($\mathbf{Q}$, $\mathbf{K}$) & $\checkmark$ & $\checkmark$ & $\checkmark$ \\
        Indexed KV-cache write & $\checkmark$ & $\checkmark$ & $\checkmark$ \\
        Standalone SwiGLU & $\checkmark$ & $\checkmark$ & $\checkmark$ \\
        \bottomrule
    \end{tabular}

\end{table}

\begin{figure*}[!htbp]
    \centering
    \includegraphics[width=\textwidth]{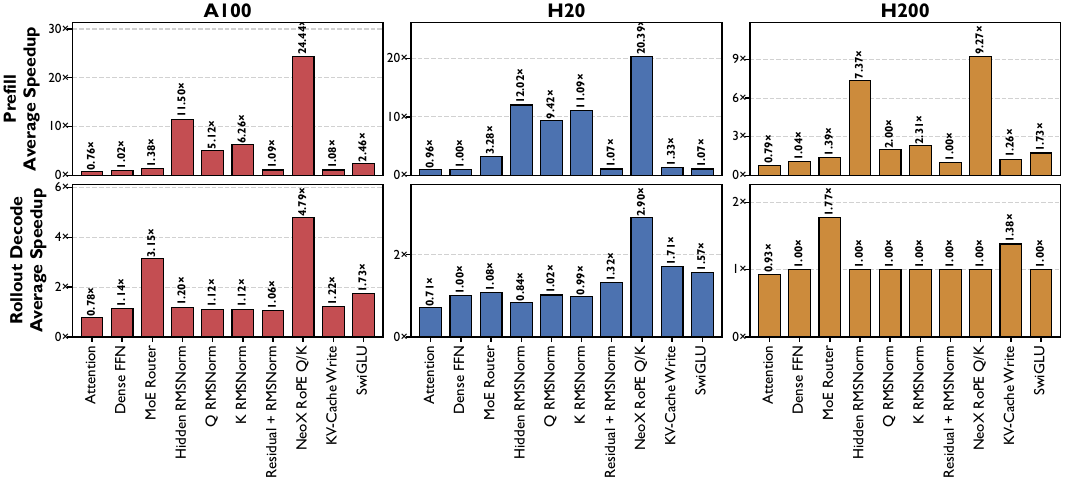}

    \caption{Operator speedups across prefill and rollout-decode workloads on A100, H20, and H200. Attention bars show average speedups of GPT-6-optimized kernels. Panels use independent vertical scales; operator labels apply to both rows.}
    \label{fig:operator-speedup}
    
\end{figure*}

\noindent\textbf{Operator performance.} Attention's average prefill and decode speedups remain below one on all three GPUs (\autoref{fig:operator-speedup}). Hidden RMSNorm and NeoX RoPE average $10.30\times$ and $18.03\times$ prefill speedups; NeoX RoPE averages $2.90\times$ for decode. Layer-level gains (\autoref{fig:per-layer-execution-times}) combine avoided policy-update recomputation with other operators' improvements, whose contributions depend on model-specific tensor dimensions and workloads.

\noindent\textbf{Attention-development time.} GPT-6 generates attention source-code modifications. Recorded optimization spans 2.51 hours on H200 and 1.38 hours on A100, including setup and between-command intervals. Across both GPUs, 113 compilation, correctness-validation, and performance-measurement jobs average 45.0 seconds of execution, including rejected-candidate jobs and 14 unsuccessful jobs.

\FloatBarrier
\subsection{Q4: Kernel-Search Comparison}
\label{subsec:openevolve-comparison}

We compare \sys with OpenEvolve~\cite{sharma2025openevolve}, inspired by AlphaEvolve~\cite{novikov2025alphaevolve}, and SkyDiscover~\cite{liu2026skydiscover} on unified attention using the workloads in \S\ref{subsec:breakdown}. All systems share a starting implementation $\mathcal{I}_0$ verified under $\mathcal{S}$, editable source-code regions, code-generating model, sampling settings, compiler and profiler interfaces, and evaluator. Searches use equal optimization-step counts rather than independent stopping; LLM-token usage is measured, not matched. Baselines retain native prompts and search organizations; \sys selects modifications and manages branches through persistent IR records. Workloads have unit weights in Equation~\eqref{eq:joint-performance} and latency limits of $1.1\times$ their fixed baselines in $\mathcal{S}$. Best aggregate speedup, $T(\mathcal{I}_0)/T(\mathcal{I}^{*})$, divides the starting implementation's summed workload latency by the lowest sum achieved by a verified candidate satisfying every limit; compilation or correctness failures cannot improve it. We count all model input and output tokens, including unsuccessful modifications and retries. Metrics are arithmetic means of five independent trials.

\begin{figure}[!htbp]
    \centering
    \includegraphics[width=0.618\textwidth]{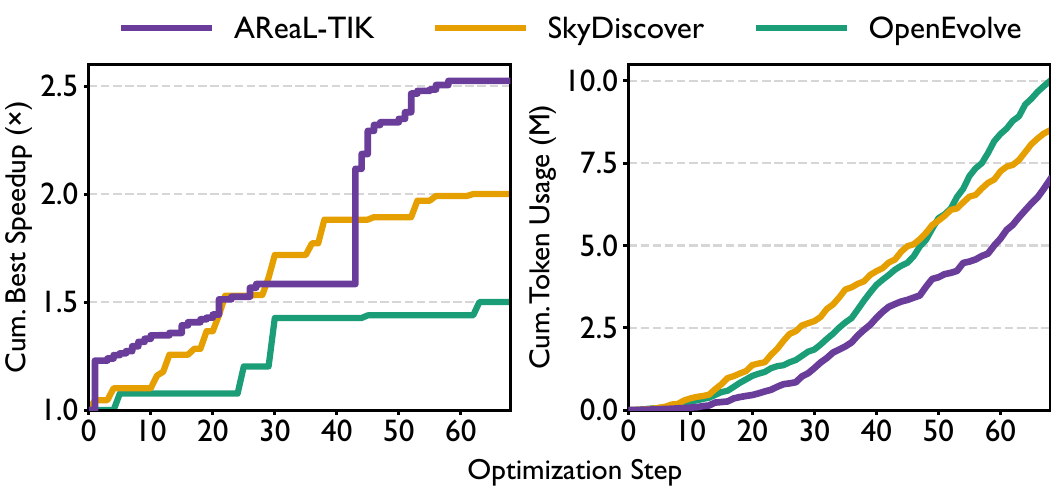}

    \caption{H20 unified-attention search versus optimization step: best aggregate speedup over the shared bitwise-consistent starting implementation (left) and cumulative LLM tokens (right), averaged over five independent trials.}
    \label{fig:openevolve-comparison}
    
\end{figure}

\noindent\textbf{Search results.} Best aggregate speedups reach $2.52\times$ for \sys, $2.00\times$ for SkyDiscover, and $1.50\times$ for OpenEvolve, largely stabilizing after approximately 65 steps (\autoref{fig:openevolve-comparison}). All retain program history and evaluation feedback, so this compares complete search procedures, not the IR alone. \sys uses $7$M tokens versus SkyDiscover's $8.5$M and OpenEvolve's $10$M, with lower cumulative usage throughout optimization. For speedup $S$ and cumulative tokens $N$, efficiency $(S-1)/(N/10^6)$ measures improvement over the starting speedup of one per million tokens. At step 60, \sys reaches $2.52\times$ using $5.21$M tokens versus OpenEvolve's $1.44\times$ using $8.73$M, a $5.8\times$ efficiency advantage.

\FloatBarrier
\subsection{Q5: Ablation Study}
\label{subsec:ablation}

Ablations use Q4's unified-attention setup, workloads, optimization-step count, and evaluator, reporting arithmetic means of five independent trials. Mandatory correctness checks, unit workload weights, and $1.1\times$ per-workload latency limits remain unchanged; only admissible candidates can improve results. We evaluate:

\begin{itemize}[leftmargin=*]
    \item \textit{Without persistent IR history.} Only the current-best implementation and its latest audit record are available; earlier attempts' records are withheld.
    \item \textit{Without explicit numerical requirements.} Requirements between entry points are withheld during modification proposals but still enforced by the evaluator.
    \item \textit{Without detailed audit evidence.} Only correctness outcomes and aggregate execution time are provided; per-workload latency, dispatch, compiler-resource, profiler, and generated-instruction evidence are withheld.
    \item \textit{Without optimization-opportunity records.} Modifications are proposed directly from the remaining state, without selecting evidence-grounded opportunities from $\mathcal{O}_t$.
    \item \textit{Without multi-branch lineage management.} Each modification starts from the current best; implementations that do not replace it cannot be continued, reactivated, or merged.
\end{itemize}

\begin{figure}[!htbp]
    \centering
    \includegraphics[width=0.618\textwidth]{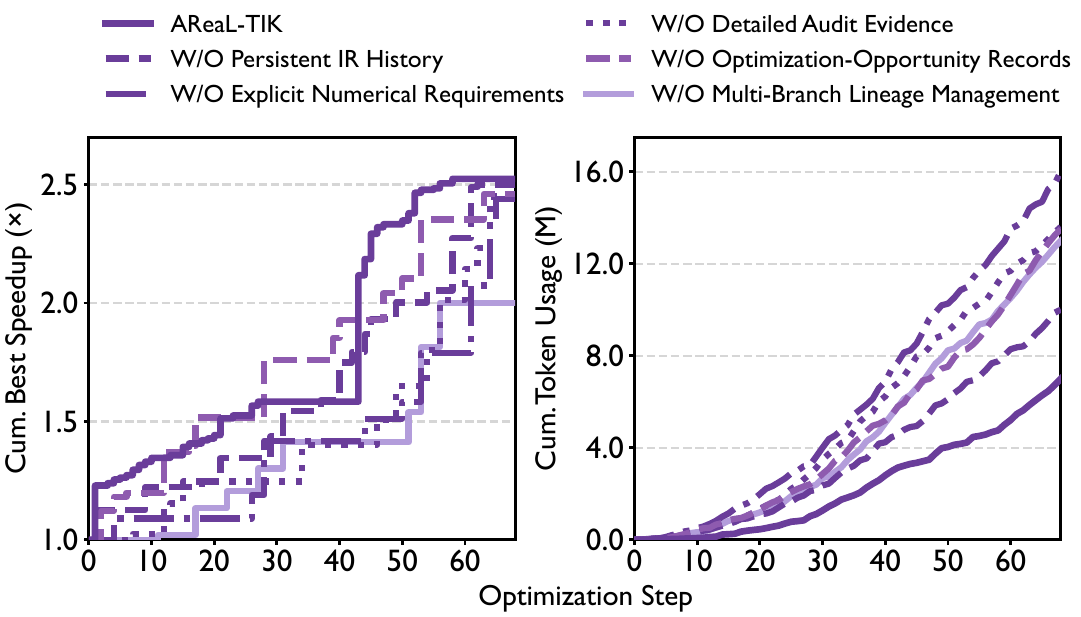}

    \caption{Ablation study on unified-attention optimization. Best aggregate speedup (left) and cumulative LLM-token usage (right), averaged across five independent trials and plotted against optimization step.}
    \label{fig:ablation-curves}
    
\end{figure}

\noindent\textbf{Ablation results.} Removing lineage management reduces speedup most: $2.00\times$ versus the complete system's $2.52\times$ (\autoref{fig:ablation-curves}). Removing persistent history yields $2.50\times$; removing opportunity records, detailed audit evidence, or numerical requirements yields $2.44$--$2.46\times$. Every ablation exceeds the complete system's $7$M tokens. Withholding numerical requirements consumes the most at $16$M, despite unchanged correctness enforcement. The opportunity-record, audit-evidence, history, and lineage ablations consume $13.5$M, $13.6$M, $10$M, and $13$M, respectively. Within the step budget, history mainly affects token use; lineage management affects both token use and attained speedup.

\begin{samepage}
\section{Conclusion}

\sys uses an optimization IR as persistent memory for joint source-code search across rollout and policy-update kernels. Numerical requirements, workload evidence, and verified alternatives support multi-step exploration under bitwise-consistency constraints. Unified kernels permit rollout-log-probability reuse when the policy snapshot and probability processing match the objective. End-to-end evaluation yields $1.10\times$ average throughput relative to AReaL with recomputation, with comparable average training reward but configuration-dependent gains and regressions. Isolated-layer profiling yields $1.40\times$ average speedup in summed phase time. Operator-level evaluation covers 10 operators across three GPU architectures, passing the prescribed bitwise checks.

\end{samepage}

\bibliographystyle{unsrt}
\bibliography{references}

\clearpage
\appendix
\section{Training Hyperparameters}
\label{app:training-hyperparameters}

\begin{longtable}{@{}>{\centering\arraybackslash}p{0.14\textwidth}|p{0.23\textwidth}|p{\dimexpr0.63\textwidth-4\tabcolsep-2\arrayrulewidth\relax}@{}}
    \caption{Training Hyperparameters for Qwen2.5-1.5B and Qwen3-4B on GSM8K. Unless noted, settings apply to both systems, both algorithms, and all data-staleness values.}\label{tab:training-hyperparameters}\\
        \toprule
        \textbf{Category} & \textbf{Configuration} & \textbf{Value} \\
        \midrule
        \multirow{4}{*}{Workload}
            & Models & \texttt{Qwen/Qwen2.5-1.5B-Instruct}; \texttt{Qwen/Qwen3-4B} \\
            & Dataset & GSM8K \\
            & Algorithms & PPO with a trained critic; GRPO without a critic \\
            & Data staleness value & $s\in\{0,2,4\}$ \\
        \midrule

        \multirow{6}{*}{Data}
            & Seeds & 1, 42, 1023 \\
            & Training batch & 64 prompts \\
            & Validation batch & 64 prompts; no shuffle or drop-last; four workers \\
            & Responses per prompt & 4 (256 responses per update) \\
            & Prompt length limit & 1,024 tokens \\
            & Decoding length limit & 1,024 tokens \\
        \midrule

        \multirow{5}{*}{Execution}
            & Hardware & $8\times$ H20 \\
            & Rollout backend & \texttt{sglang:d4p1t1} \\
            & Policy-update backend & \texttt{fsdp:d4p1t1} for the policy model (actor), KL-reference model, and PPO critic \\
            & Precision & \texttt{BF16} parameters; \texttt{FP32} actor-gradient reduction and optimizer state \\
            & Rollout concurrency & 256 maximum concurrent rollouts \\
        \midrule

        \multirow{9}{*}{Optimization}
            & Optimizer & Adam \\
            & Learning rate & $1\!\times\!10^{-6}$ \\
            & Weight decay & 0.017 \\
            & Momentum coefficients & $\beta_1=0.9$; $\beta_2=0.999$ \\
            & Adam epsilon & $\epsilon=10^{-8}$ \\
            & Learning-rate schedule & Cosine \\
            & Warmup & 5\% of training \\
            & Minimum learning rate & 0.1 times the configured learning rate \\
            & Gradient clipping & 0.5 \\
        \midrule

        \multirow{6}{*}{Objective}
            & Actor minibatches & 4 \\
            & Clipping coefficient & 0.2 \\
            & Reward processing & Scale 1; bias 0; clip 20 \\
            & Discount & 1 \\
            & Advantage normalization & Batch mean and standard deviation \\
            & KL regularization & Coefficient 0.001; \texttt{k3} estimator \\
        \midrule

        \multirow{2}{*}{Microbatch}
            & Actor token cap & 10,240 for Qwen2.5-1.5B GRPO; 2,048 otherwise \\
            & KL-reference token cap & 10,240 \\
        \midrule

        PPO & Generalized advantage estimation & $\lambda=1$ \\
        \midrule

        \multirow{4}{*}{GRPO}
            & Reward-normalization scope & Responses to the same prompt \\
            & Group size & 4 responses per prompt \\
            & Standard-deviation estimator & Unbiased \\
            & Normalization epsilon & $10^{-5}$ \\
        \bottomrule
\end{longtable}

\end{document}